\RequirePackage[running]{lineno}

\documentclass[aps,prl,nofootinbib,showpacs,superscriptaddress,letterpaper,amsmath,amssymb]{revtex4}

\usepackage{graphicx}  
\usepackage{dcolumn}   
\usepackage{bm}        
\usepackage{amssymb}   
\usepackage{feynmf}
\usepackage{slashed}
\usepackage{multirow}
\usepackage{hyperref}
\usepackage{xcolor}
\hypersetup{
    colorlinks = true,
    citecolor = red,
}
\def\gsim{\lower0.5ex\hbox{$\:\buildrel >\over\sim\:$}}
\def\lsim{\lower0.5ex\hbox{$\:\buildrel <\over\sim\:$}}

\newcommand{\be}{\begin{equation}}
\newcommand{\ee}{\end{equation}}
\newcommand{\bea}{\begin{eqnarray}}
\newcommand{\eea}{\end{eqnarray}}

\newcommand{\nbox}{{\,\lower0.9pt\vbox{\hrule \hbox{\vrule height 0.2 cm
\hskip 0.2 cm \vrule height 0.2 cm}\hrule}\,}}

\def\missET {{\not\!\! E_\mathrm{T}}}

\begin{document}

\thispagestyle{empty}
\vspace*{-3.5cm}

\vspace{0.5in}

\title{Sensitivity of future collider facilities to WIMP pair production via effective operators and light mediators}

\begin{center}
\begin{abstract}
We present extrapolations of the current mono-jet searches at the LHC to potential future hadron collider facilities: LHC14, as well as $pp$ colliders with $\sqrt{s}=33$ or 100 TeV.   We consider both the effective operator approach as well as one example of a light mediating particle.
\end{abstract}
\end{center}

\author{Ning Zhou}
\affiliation{Department of Physics and Astronomy, University of California, Irvine, CA 92697}
\author{David Berge}
\affiliation{GRAPPA Institute, University of Amsterdam, Netherlands}
\author{LianTao Wang}
\affiliation{Department of Physics, University of Chicago, Chicago, IL}
\author{Daniel Whiteson}
\author{Tim Tait}
\affiliation{Department of Physics and Astronomy, University of
  California, Irvine, CA 92697}

\pacs{}
\maketitle


Though the presence of dark matter in the universe has been
well-established, little is known of its particle nature or its
non-gravitational interactions.  A vibrant experimental program
is searching for a weakly interacting massive particle (WIMP), denoted as
$\chi$, and interactions with standard model particles via some
as-yet-unknown mediator.  

\begin{figure}
\includegraphics[width=0.3\linewidth]{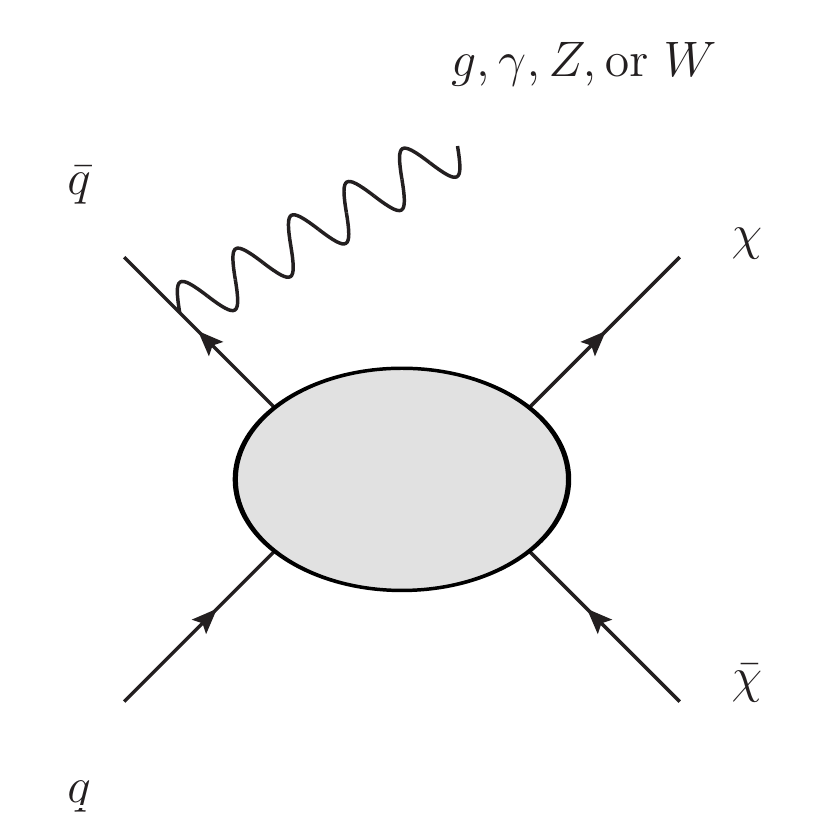}
\caption{ Pair production of WIMPs ($\chi\bar{\chi}$) in proton-proton
  collisions at the LHC via an unknown intermediate state, with initial-state radiation of a standard
  model particle.}
\label{fig:diag}
\end{figure}

One critical component of this program is the search for
pair-production of WIMPs at particle colliders, specifically
$pp\rightarrow \chi\bar{\chi}$ at the LHC via some unknown
intermediate state.  If the mediator is too heavy to be
resolved, the interaction can be modeled as an effective field theory
with a four-point interaction, otherwise an explicit  model is needed
for the heavy mediator. As the final state WIMPs are invisible to the
detectors, the events can only be seen if there is associated
initial-state radiation of a standard model
particle~\cite{Beltran:2010ww,Fox:2011pm,Goodman:2010ku}, see Fig~\ref{fig:diag}, recoiling against the dark matter pair.

The LHC collaborations have reported limits on the cross section of
$pp\rightarrow  \chi\bar{\chi}+X$ where $X$ is a
hadronic jet~\cite{atlasjet,cmsjet}, photon~\cite{atlasphoton,cmsphoton}, and
other searches have been repurposed to study the cases where $X$ is a
$W$~\cite{monow} or $Z$ boson~\cite{atlaszz,monoz}. In each case,
limits are reported in terms of the mass scale
$M_\star$ of the unknown interaction expressed in an effective field
theory~\cite{Beltran:2008xg,Beltran:2010ww, Fox:2011pm,Goodman:2010ku,
  Shepherd:2009sa,Cao:2009uw,Goodman:2010yf,Bai:2010hh,Rajaraman:2011wf,Cotta:2012nj,Petriello:2008pu,Gershtein:2008bf},
though the limits from the mono-jet mode are the most powerful~\cite{dmcombo}.

In this paper,  we study the sensitivity of possible future
proton-proton colliders in various configurations (see
Table~\ref{tab:fac}) to WIMP pair production using the mono-jet final
state. We consider both effective operators and one example of a real, heavy
$Z'$-boson mediator.

\section{Analysis Technique}

The analysis of jet$+\missET$ events uses a sample of events with one
or two high $p_\mathrm{T}$ jets and large $\missET$, with angular cuts
to suppress events with two back-to-back jets (multi-jet
background). The dominant remaining background is $Z\rightarrow
\nu\bar{\nu}$ in association with jets, which is indistinguishable
from the signal process of $\chi\bar{\chi}$+jets.

The estimation of the background at large $\missET$ Is problematic in
simulated samples, due to the difficulties of accurately modeling the
many sources of $\missET$.  The experimental results, therefore, rely
on data-driven background estimates, typically extrapolating the
$Z\rightarrow \nu\bar{\nu}$ contribution from $Z\rightarrow\mu\mu$
events with large $Z$ boson $p_\mathrm{T}$.

In this study, we begin from experimentally reported
values~\cite{atlasjet,cmsjet} of the background estimates and signal
efficiencies (at $\sqrt{s}=7$ TeV, $\mathcal{L}=5$ fb$^{-1}$, $\missET
> 350$ GeV), and use simulated samples to extrapolate to higher
center-of-mass energies, where no data is currently available.  At the
higher collision energies and instantaneous luminosities of the
proposed facilities, the rate of multi-jet production will also be
higher, requiring higher $\missET$ thresholds to cope with the
background levels and the trigger rates.

\subsection{Simulated Samples}

We generate signal events as well as events for the dominant
background $Z\rightarrow \nu\bar{\nu}$+jets in {\sc
  madgraph}5~\cite{MG}, with showering and hadronization by {\sc
  pythia}~\cite{PY}. In each case, we generate events with zero or one
hard additional colored parton and use the MLM scheme to match the
matrix-element calculations of {\sc madgraph}5 to the parton shower
evolution of {\sc pythia}.

\subsection{Extrapolation}

In the case of each potential facility, we must choose a $\missET$
threshold for the analysis. For a given threshold at a specific
facility, estimating the sensitivity of the jet+$\missET$ analysis
requires

\begin{itemize}
\item the dark-matter signal efficiency
\item an estimate of the $Z\rightarrow \nu\bar{\nu}$+jets background 
\item the uncertainty of the $Z\rightarrow \nu\bar{\nu}$+jets 
  background.
\end{itemize}

\begin{figure}[htp]
\includegraphics[width=0.8\linewidth]{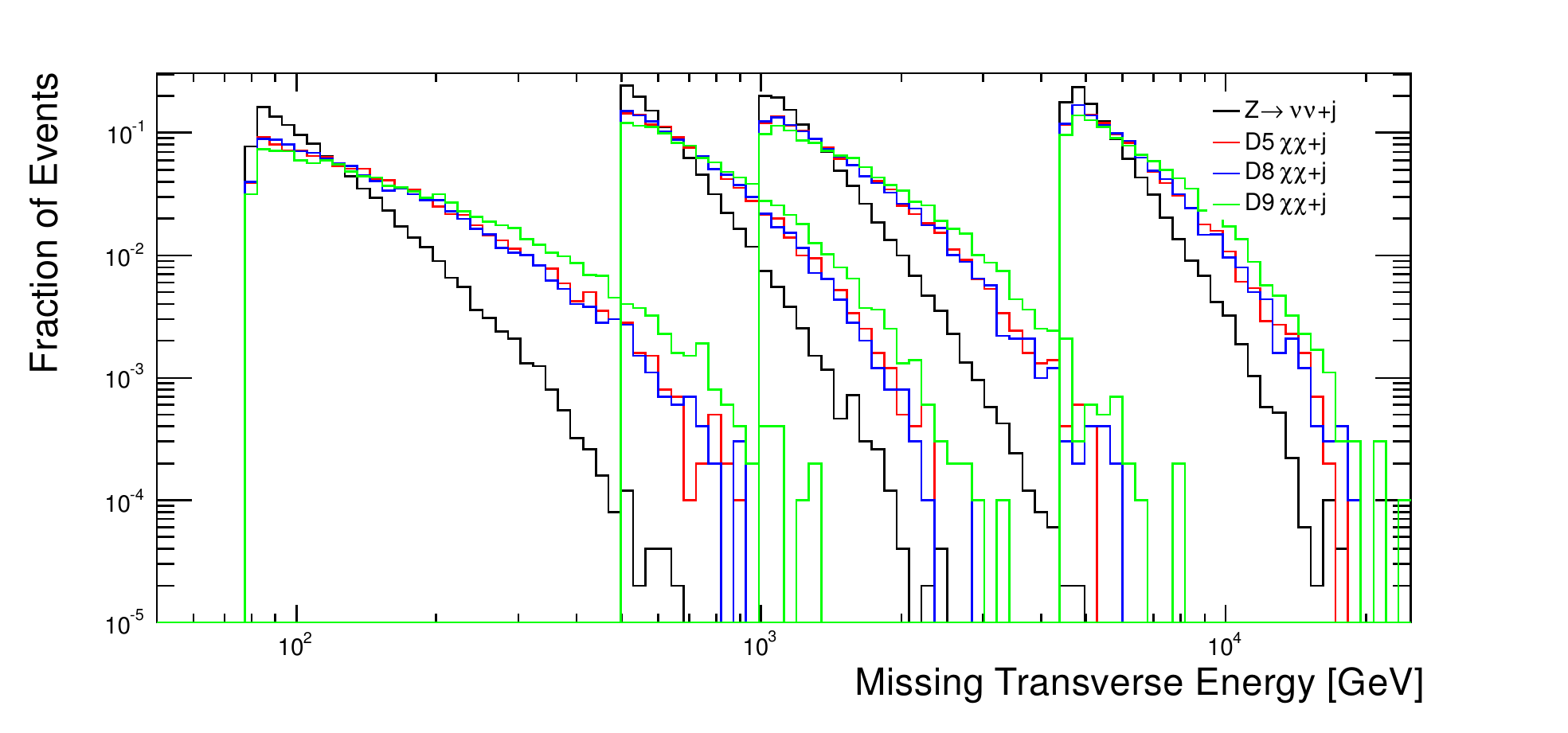}
\caption{ Missing transverse energy for signal ($\chi\bar{\chi}$+jets)
  and background ($Z\rightarrow \nu\bar{\nu}$+jets) samples generated
  at $\sqrt{s}=$7, 14, 33 and $100$ TeV with thresholds of 80, 500,
  1000 and 4500 GeV, with $m_\chi=10$ GeV.}
\label{fig:met}
\end{figure}

For the signal efficiency, we modify the reported experimental
efficiency at $\sqrt{s}=7$ TeV~\cite{atlasjet,cmsjet} to estimate the
efficiency at a higher $\missET$ cut. The estimated signal yield
$N_{\textrm{sig}}(\sqrt{s},\ \mathcal{L},\ \missET>X)$ for a facility with
center-of-mass energy $\sqrt{s}$ and integrated luminosity
$\mathcal{L}$ is

\[ N_{\textrm{sig}}(\sqrt{s},\ \mathcal{L},\ \missET>X) = \mathcal{L} \times \epsilon_0
\frac{\epsilon_{\ \missET>X}}{\epsilon_{\ \missET>350}}
\times \sigma(\sqrt{s}) \]

\noindent where $\epsilon_0$ is the published signal efficiency. In
each case, the efficiency $\epsilon_{\ \missET>X}$ of a $\missET$
threshold is measured at parton-level using the simulated samples, see
Fig.~\ref{fig:met}, and the cross sections $\sigma(\sqrt{s})$ are
leading order.

In the case of the background estimate, we extrapolate from the
reported background estimate, denoted
$N_{\textrm{bg}}^{\sqrt{s}=7,\mathcal{L}=5,\ \missET>350}$, by scaling
with an extrapolation factor $E_b$:

\[ N_{\textrm{bg}}(\sqrt{s},\ \mathcal{L},\ \missET>X) = E_b\times
N_{\textrm{bg}}^{\sqrt{s}=7,\mathcal{L}=5,\ \missET>350}\]

\noindent where $E_b$ is

\[ E_b(\sqrt{s},\ \mathcal{L},\ \missET>X) =  \frac{\mathcal{L}}{5\ \textrm{fb}^{-1}} \times \frac{\epsilon_{\
    \missET>X}}{\epsilon_{\ \missET>350}}\times
\frac{\sigma(\sqrt{s})}{\sigma(\sqrt{s}=7)} \]

\noindent
accounting for the relative efficiency of a higher $\missET$ cut and
larger background cross sections at increased center-of-mass energies.
The relative uncertainty on the background
$N_{\textrm{bg}}^{\sqrt{s}=7,\mathcal{L}=5,\ \missET>350}$ is scaled
from the reported relative uncertainty, $(\frac{\Delta
  N_{\textrm{bg}}}{N_{\textrm{bg}}})^{\sqrt{s}=7,\mathcal{L}=5,\
  \missET>350}$, using the extrapolation factor $E_b$ as given above:

\[ \frac{\Delta N_{\textrm{bg}}}{N_{\textrm{bg}}}(\sqrt{s},\ \mathcal{L},\ \missET>X) =
\frac{1}{\sqrt{E_b}} \left(\frac{\Delta
  N_{\textrm{bg}}}{N_{\textrm{bg}}}\right)^{\sqrt{s}=7,\mathcal{L}=5,\ \missET>350} \]

\begin{table}
\caption{Details of current and potential future $pp$ colliders,
  including center-of-mass energy ($\sqrt{s}$),  total integrated
  luminosity ($\mathcal{L}$), the threshold in $\missET$, and the
  estimated signal and background yields.}
\label{tab:fac}
\begin{tabular}{lrrrrrr}
\hline\hline
$\sqrt{s}$ [TeV] & $\missET$ [GeV] & $\mathcal{L}$ [fb$^{-1}$] & $N_{D5}$ & $N_{\textrm{bg}}$ \\
\hline
7 & 350 & 4.9 & 73.3 & $1970\pm 160$\\
14 & 550 & 300 & 2500 & $2200\pm 180$\\
14 & 1100 & 3000 & 3200 & $1760\pm 143$\\
33 & 2750 & 3000 & 8.2$\cdot10^4$ & $1870\pm 150$\\
100 & 5500 & 3000 & 3.4$\cdot10^6$& $2310\pm 190$\\
\hline\hline
\end{tabular}
\end{table}

Together, the background estimate with uncertainties and the
dark-matter signal efficiencies allow us to calculate the power of the
jet$+\missET$ analysis.

\section{Results for Effective Field Theories}

\begin{figure}[h]
\includegraphics[width=0.48\linewidth]{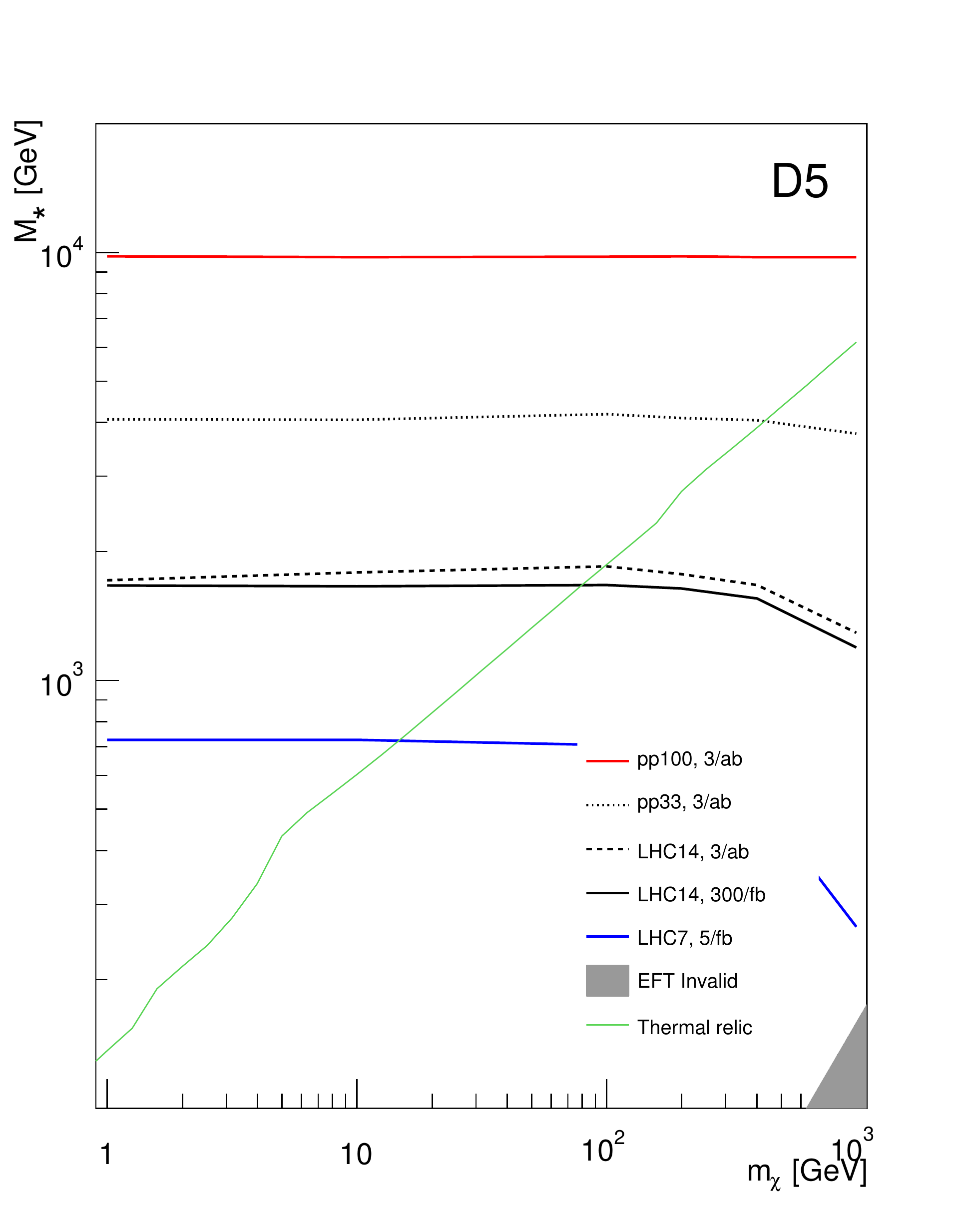}
\includegraphics[width=0.48\linewidth]{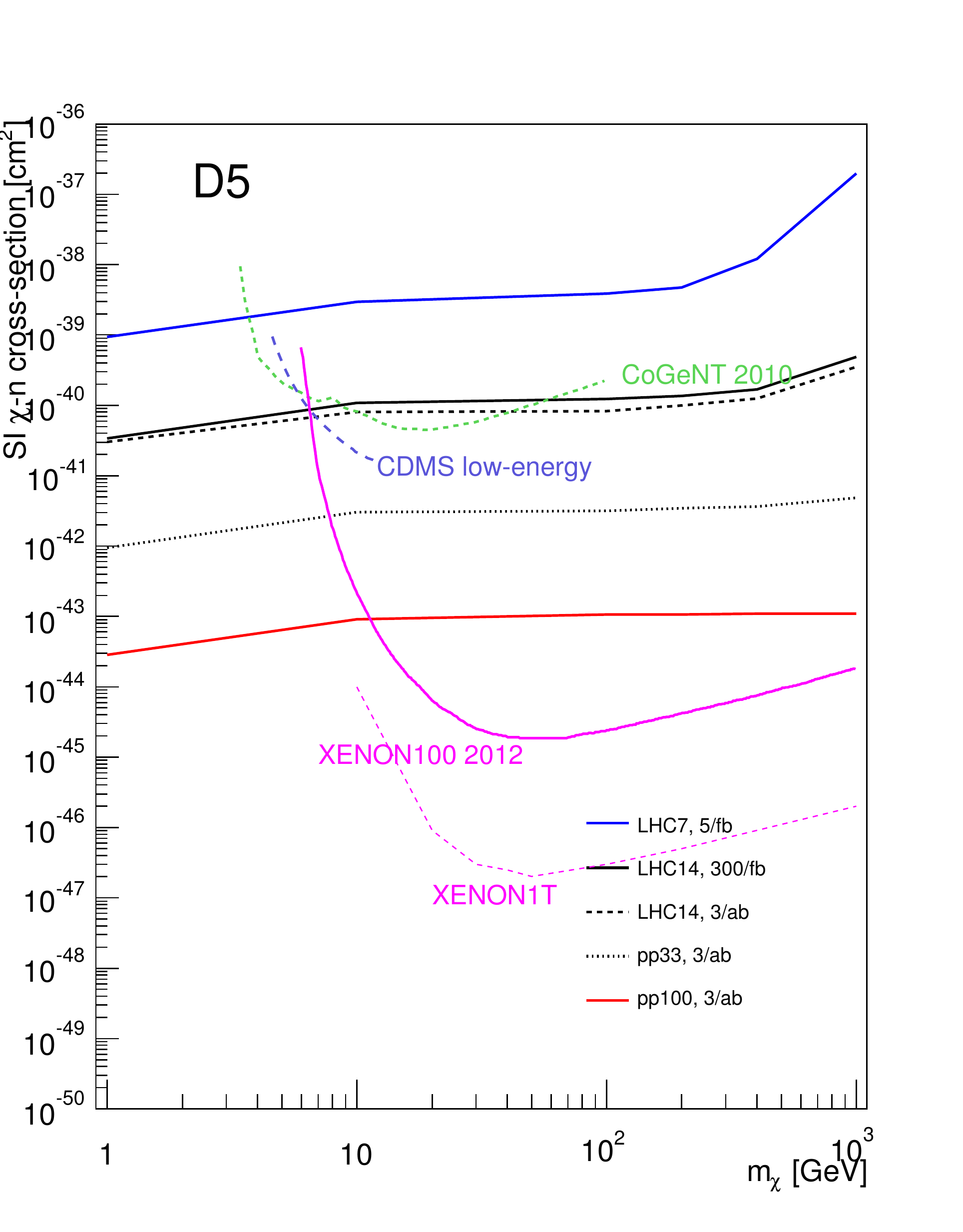}
\caption{Limits at 90\% CL in $M_\star$ (left) and in the spin-independent
  WIMP-nucleon cross section (right) for different facilities using
  the D5 operator as a function of $m_\chi$. }
\label{fig:d5}
\end{figure}

\begin{figure}[h]
\includegraphics[width=0.48\linewidth]{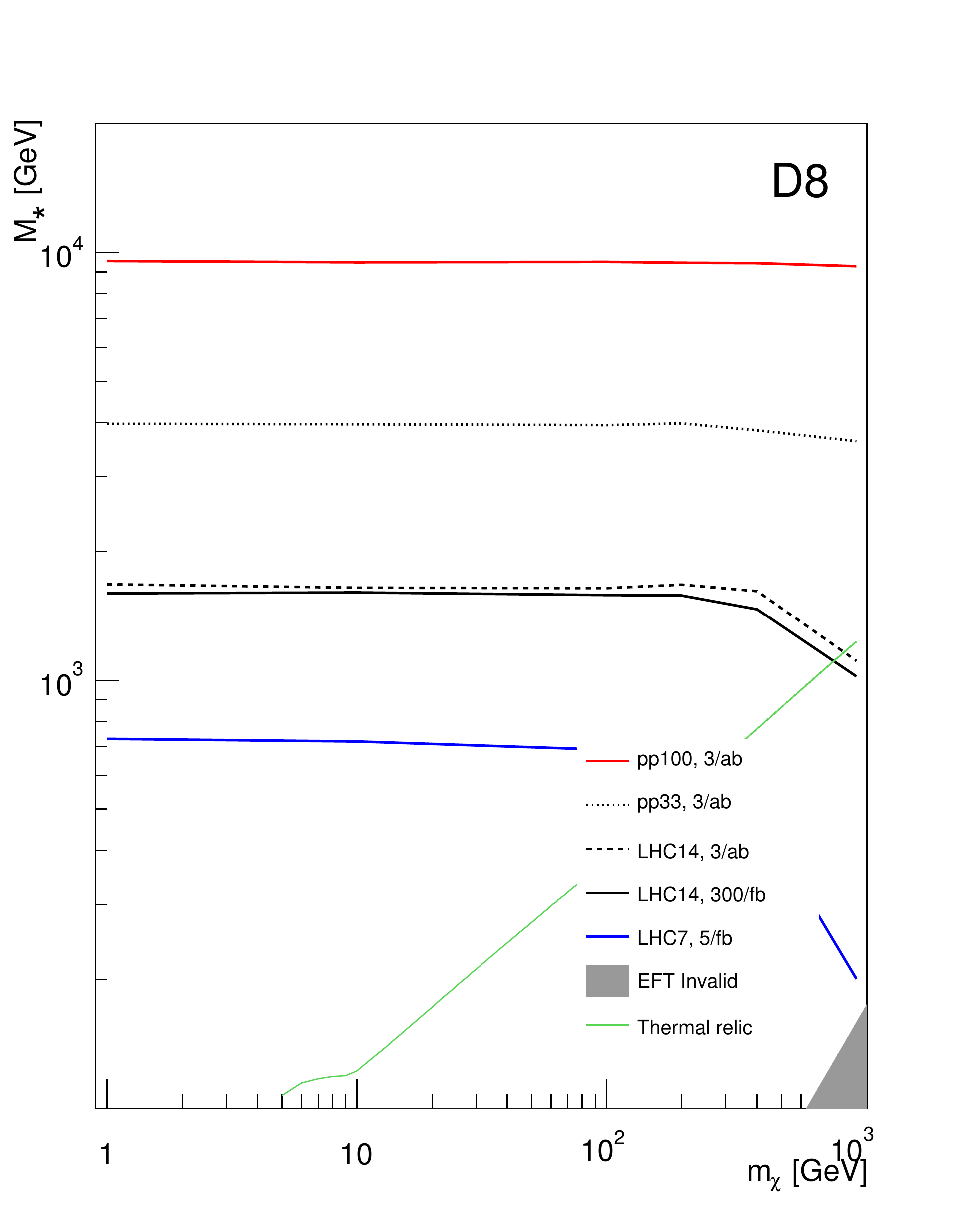}
\includegraphics[width=0.48\linewidth]{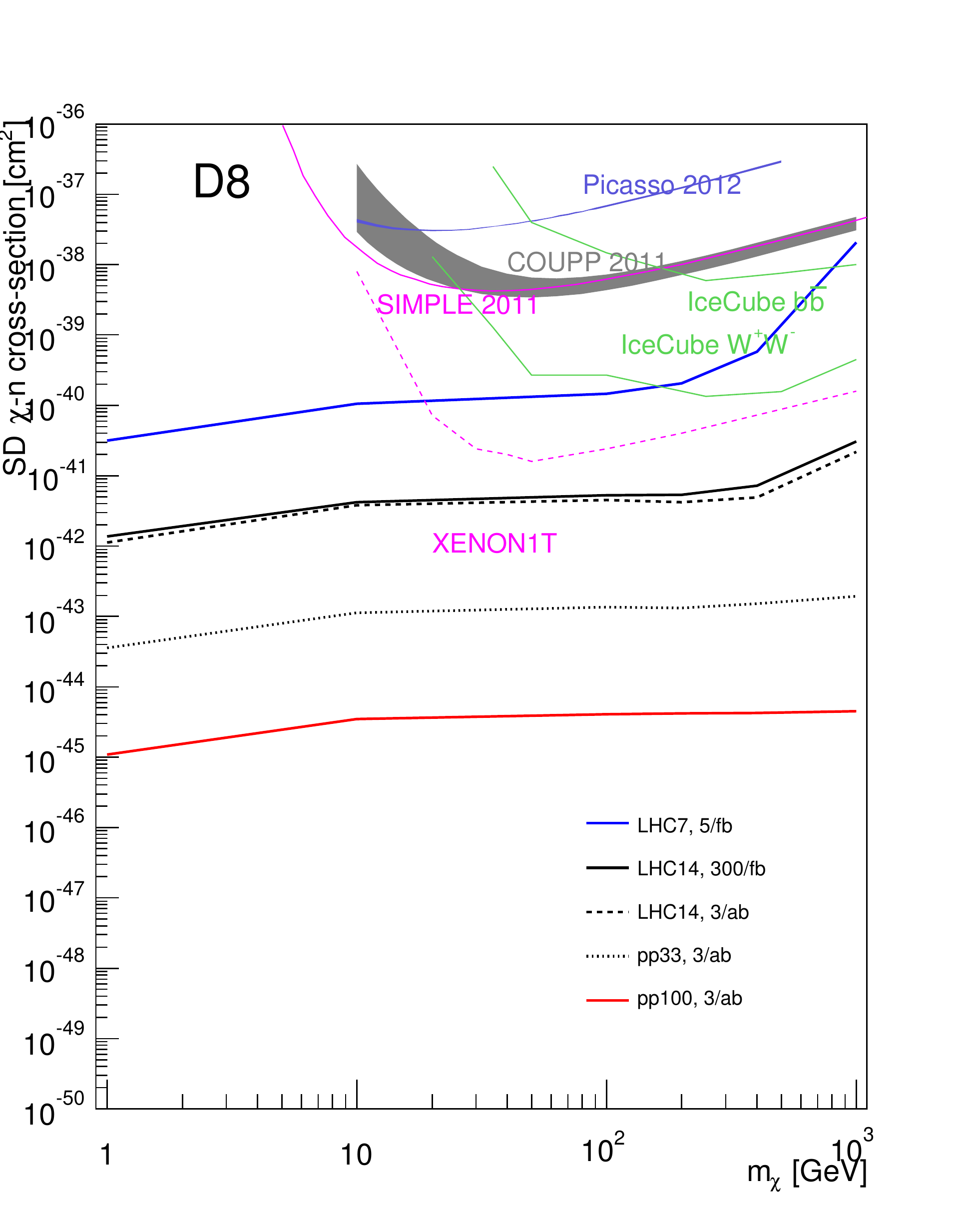}
\caption{Limits at 90\% CL in $M_\star$ (left) and in the spin-dependent
  WIMP-nucleon cross section (right) for different facilities using
  the D8 operator as a function of $m_\chi$. }
\label{fig:d8}
\end{figure}

\begin{figure}[h]
\includegraphics[width=0.48\linewidth]{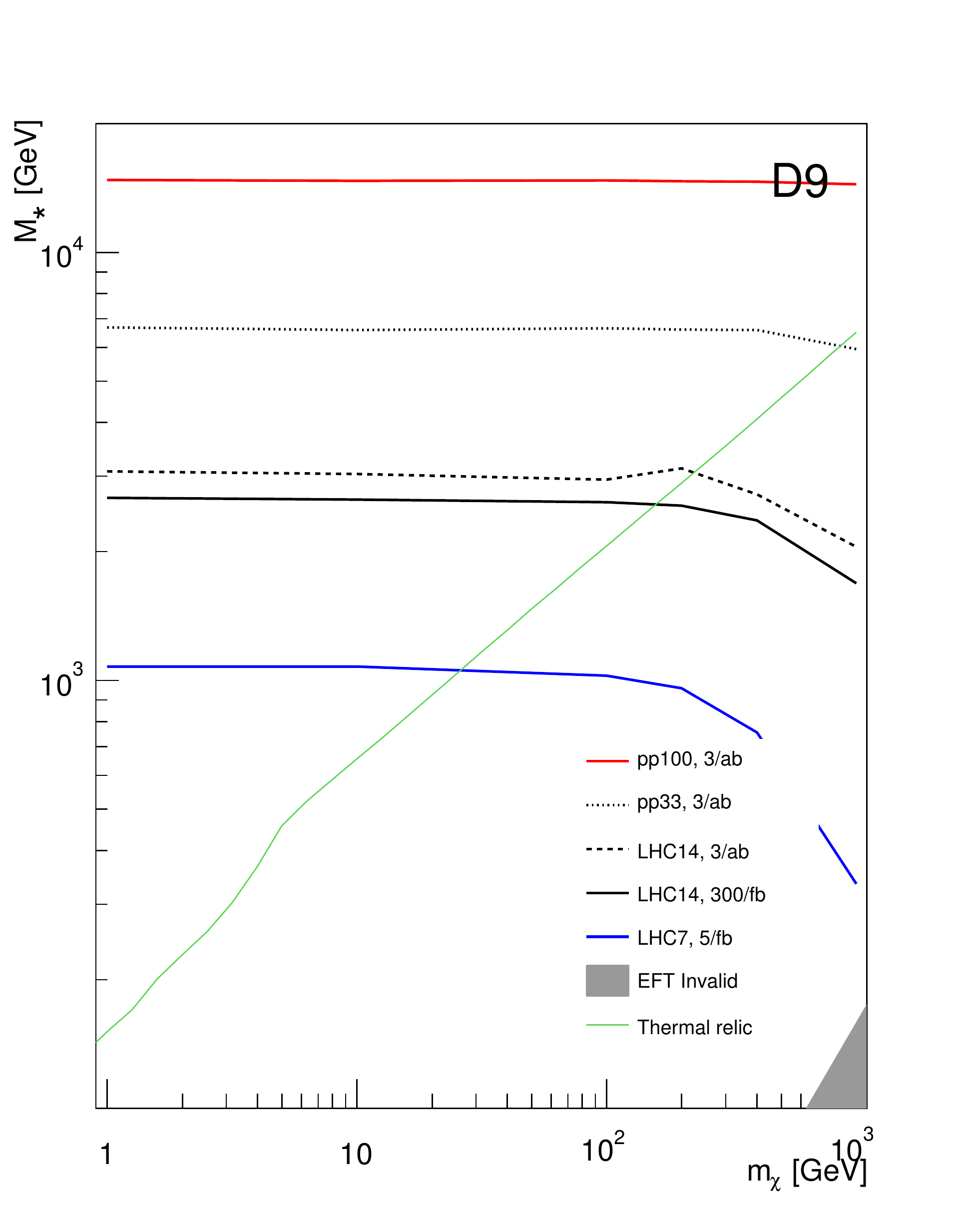}
\includegraphics[width=0.48\linewidth]{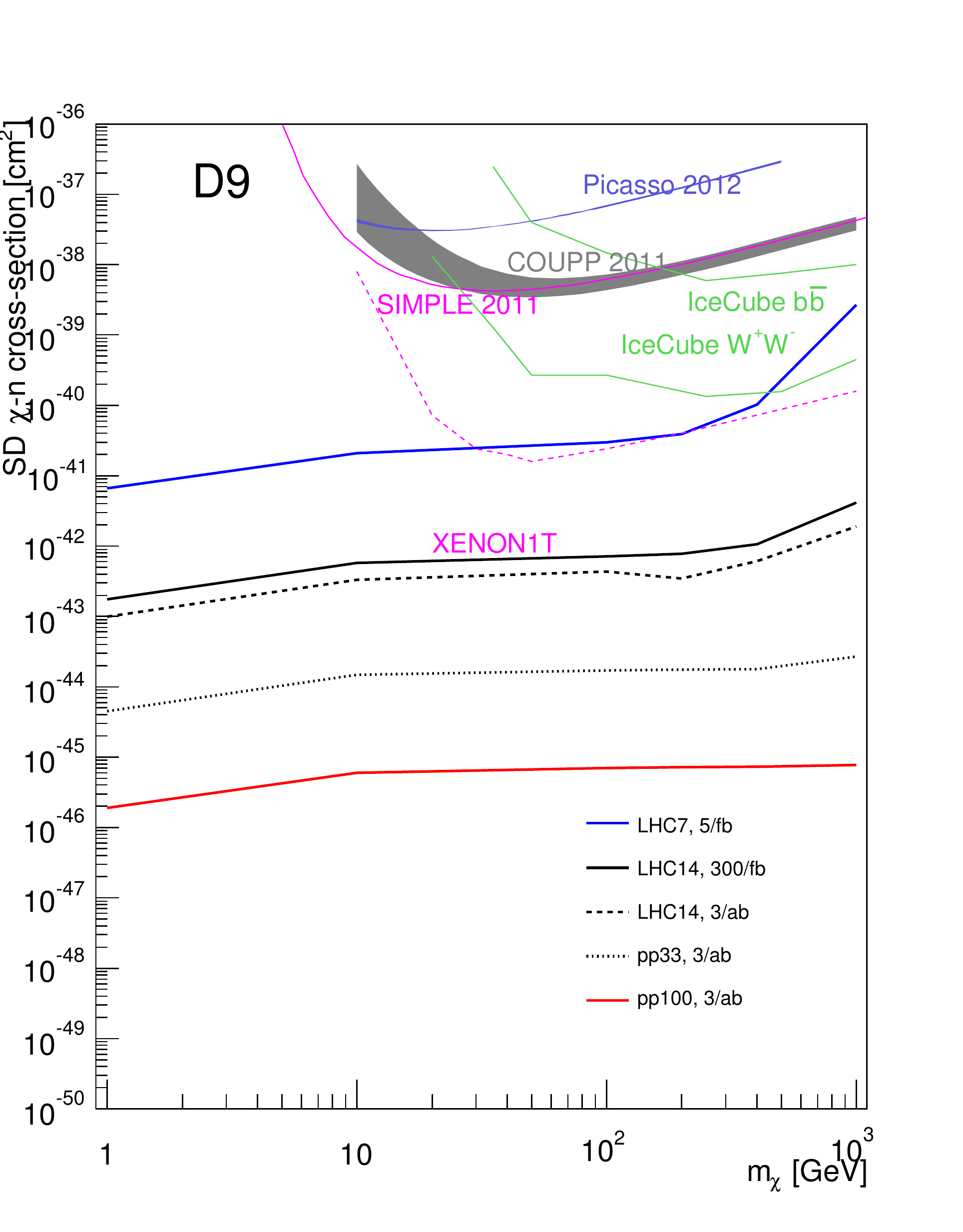}
\caption{Limits at 90\% CL in $M_\star$ (left) and in the spin-dependent
  WIMP-nucleon cross section (right) for different facilities using
  the D9 operator as a function of $m_\chi$. }
\label{fig:d9}
\end{figure}

Given the expected background and uncertainties, we use the CLs
method~\cite{cls1,cls2} to calculate expected 90\% confidence limits
on contributions from new sources. Together with the estimated signal
efficiencies, we calculate cross-section limits.  As the predicted
cross sections depend on $M_*$, we can therefore derive limits on
$M_*$, see the top panel of Figs~\ref{fig:d5},~\ref{fig:d8}, and
~\ref{fig:d9}. These are then translated in limits on the
$\chi$-nucleon cross section, see 
the right panel of Figs~\ref{fig:d5},~\ref{fig:d8}, and
~\ref{fig:d9}.

\begin{figure}[h]
\includegraphics[width=0.3\linewidth]{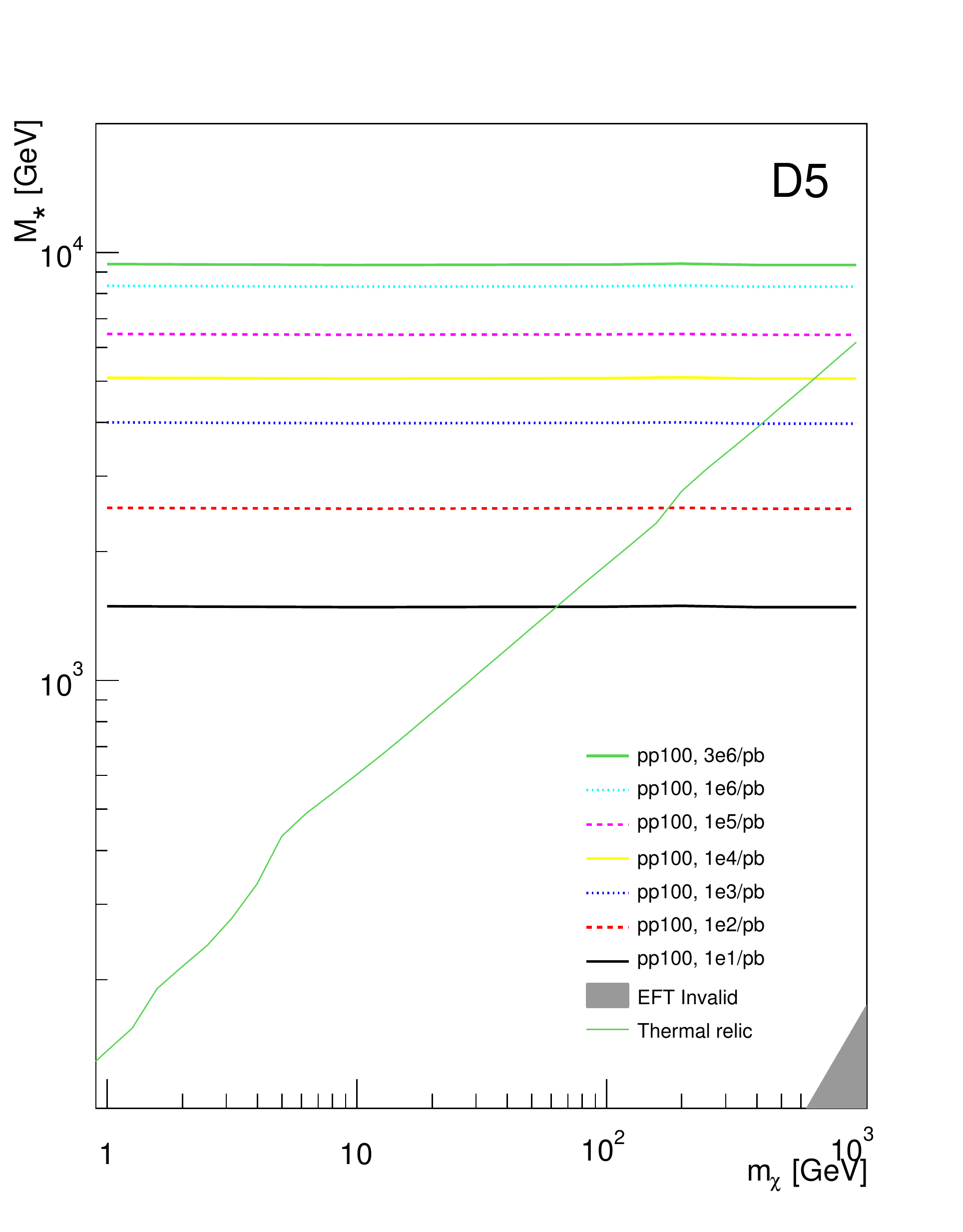}
\includegraphics[width=0.3\linewidth]{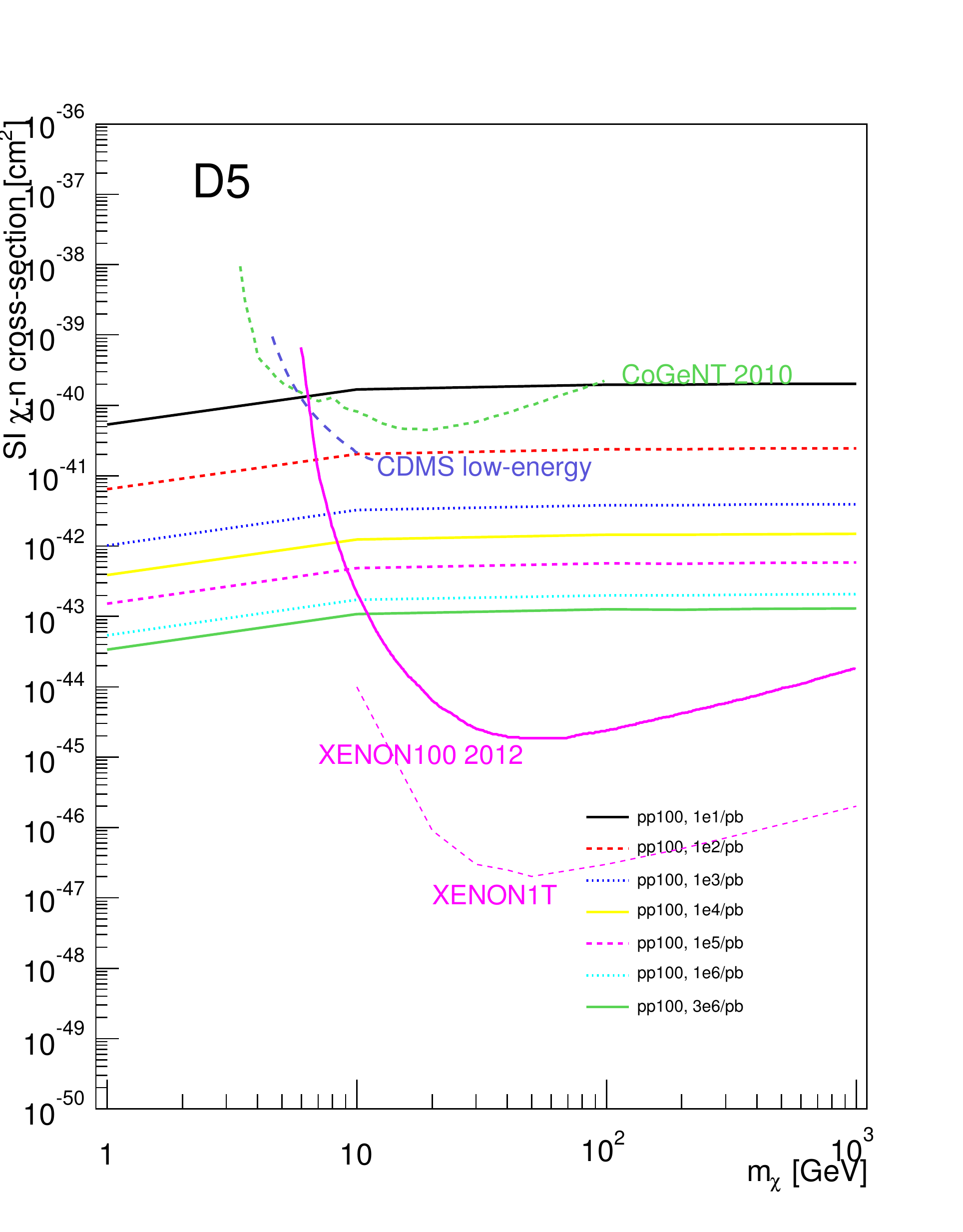}
\caption{Limits at 90\% CL in $M_\star$ (left) and in the spin-independent
  WIMP-nucleon cross section (right) for increasing luminosity using
  the D5 operator as a function of $m_\chi$. }
\label{fig:lumid5}
\end{figure}

\begin{figure}[h]
\includegraphics[width=0.3\linewidth]{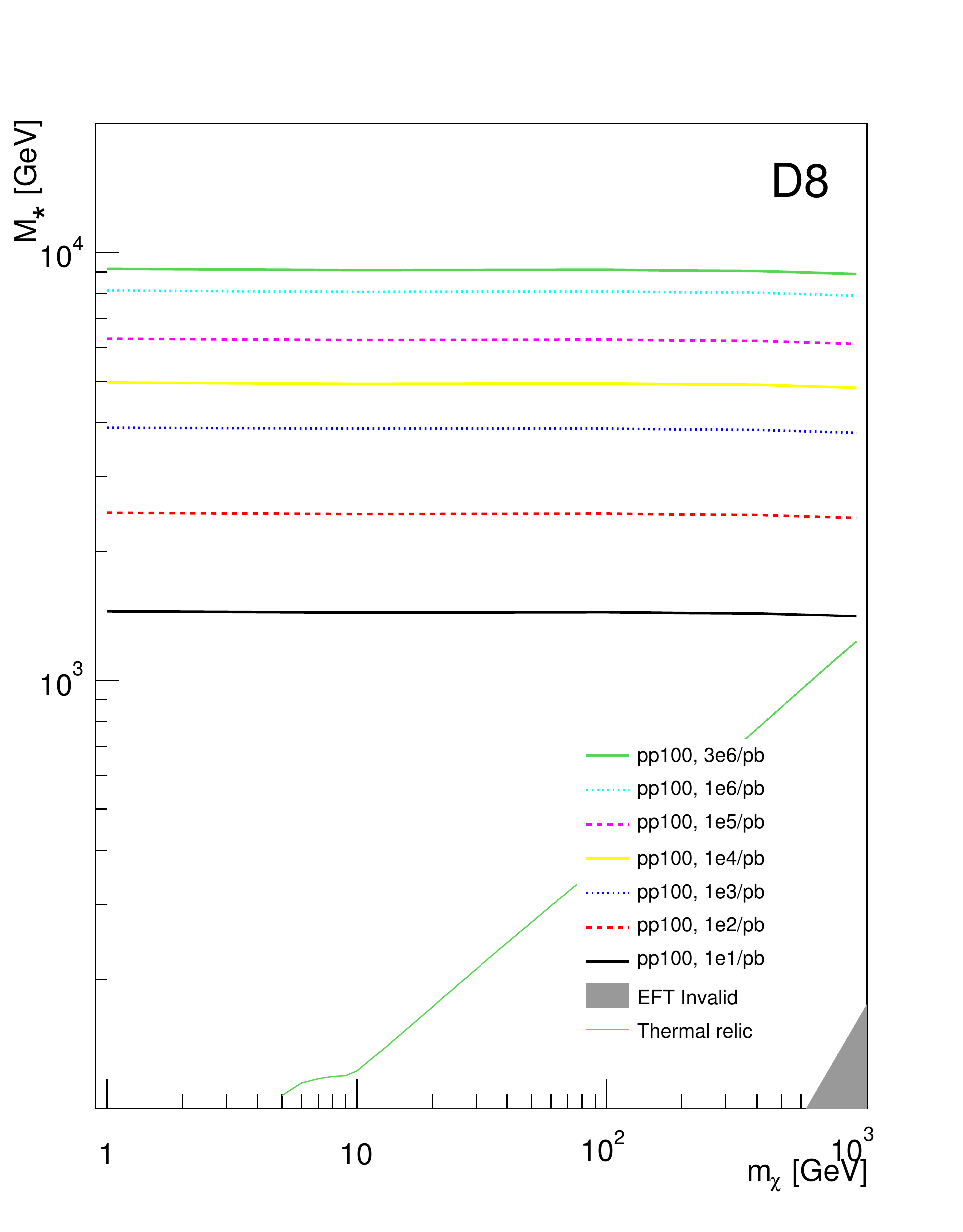}
\includegraphics[width=0.3\linewidth]{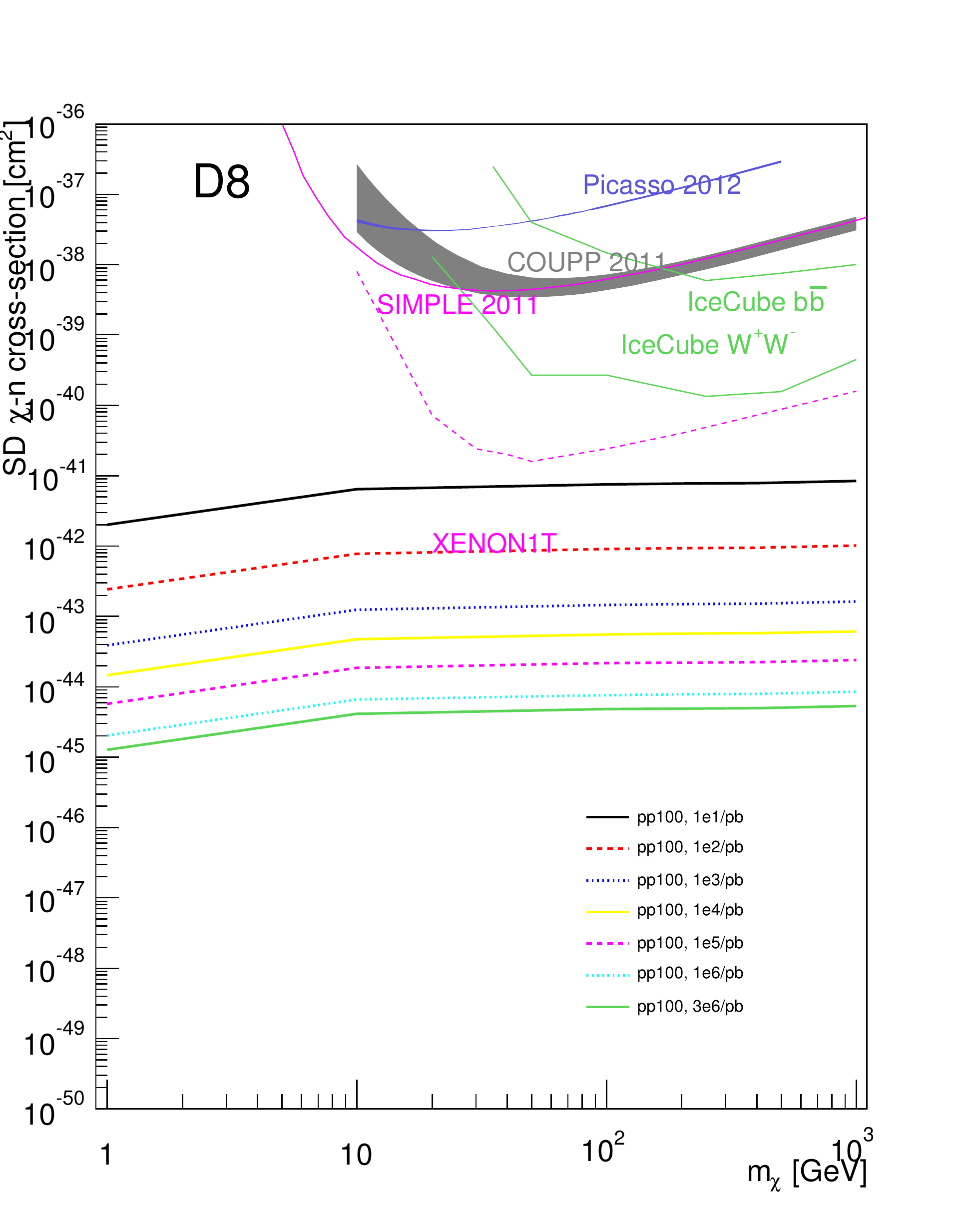}
\caption{Limits at 90\% CL in $M_\star$ (left) and in the spin-dependent
  WIMP-nucleon cross section (right) for increasing luminosity using
  the D8 operator as a function of $m_\chi$. }
\label{fig:lumid8}
\end{figure}

\begin{figure}[h]
\includegraphics[width=0.3\linewidth]{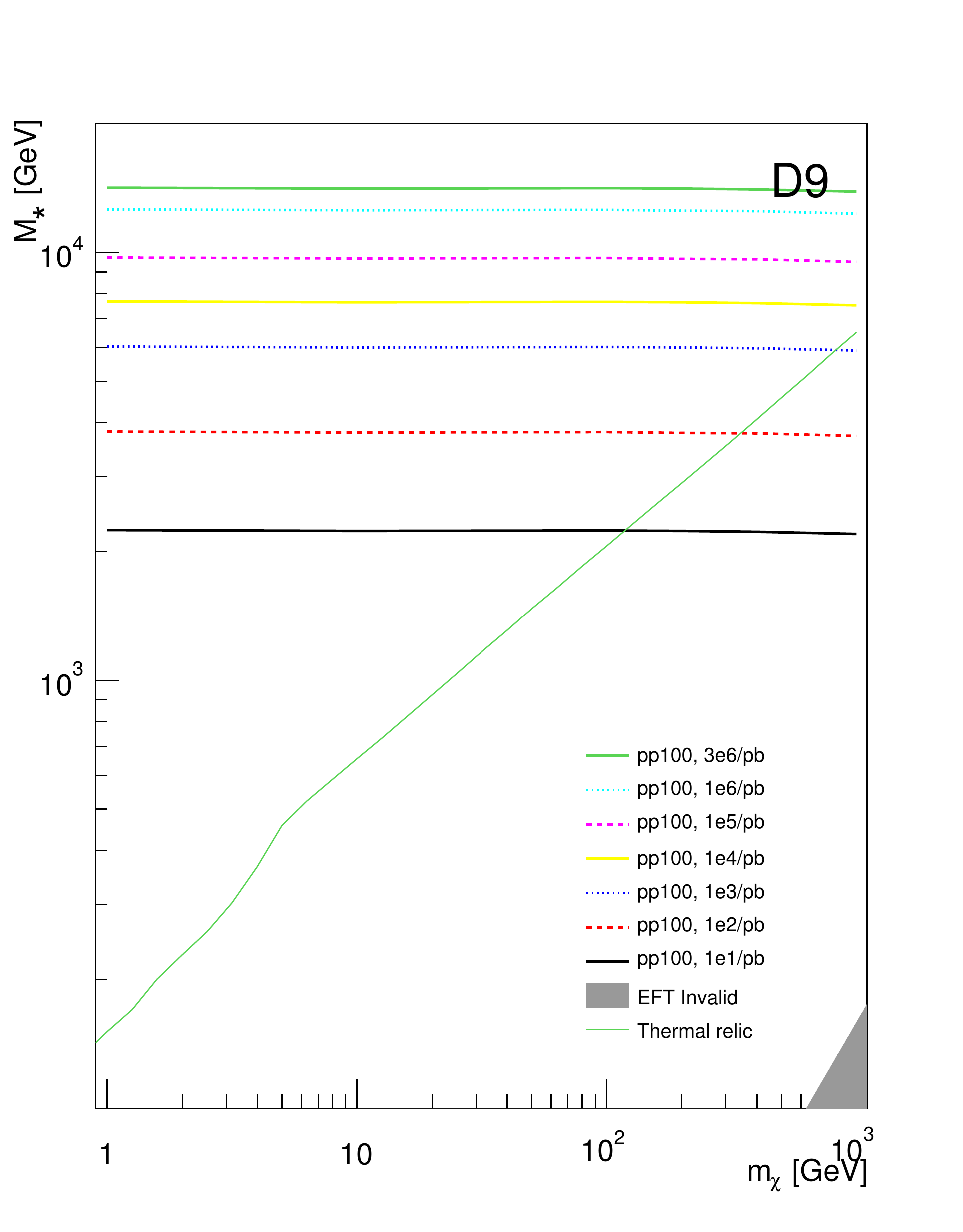}
\includegraphics[width=0.3\linewidth]{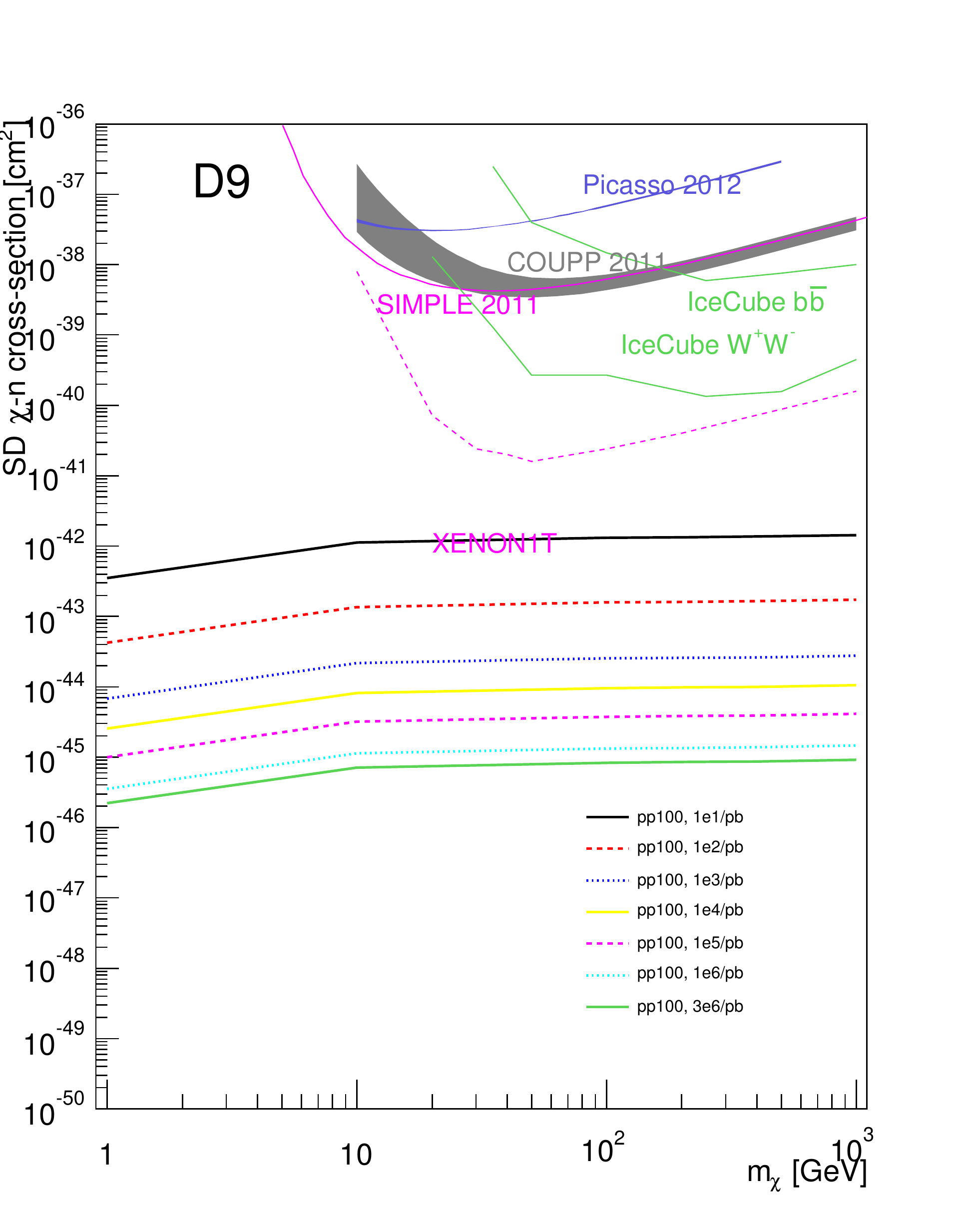}
\caption{Limits at 90\% CL in $M_\star$ (left) and in the spin-dependent
  WIMP-nucleon cross section (right) for increasing luminosity using
  the D9 operator as a function of $m_\chi$. }
\label{fig:lumid9}
\end{figure}

In addition, we study the luminosity dependence of the results at
$\sqrt{s}=100$ TeV, see Figs~\ref{fig:lumid5},~\ref{fig:lumid8}, and
~\ref{fig:lumid9}.

\begin{figure}[h]
\includegraphics[width=0.48\linewidth]{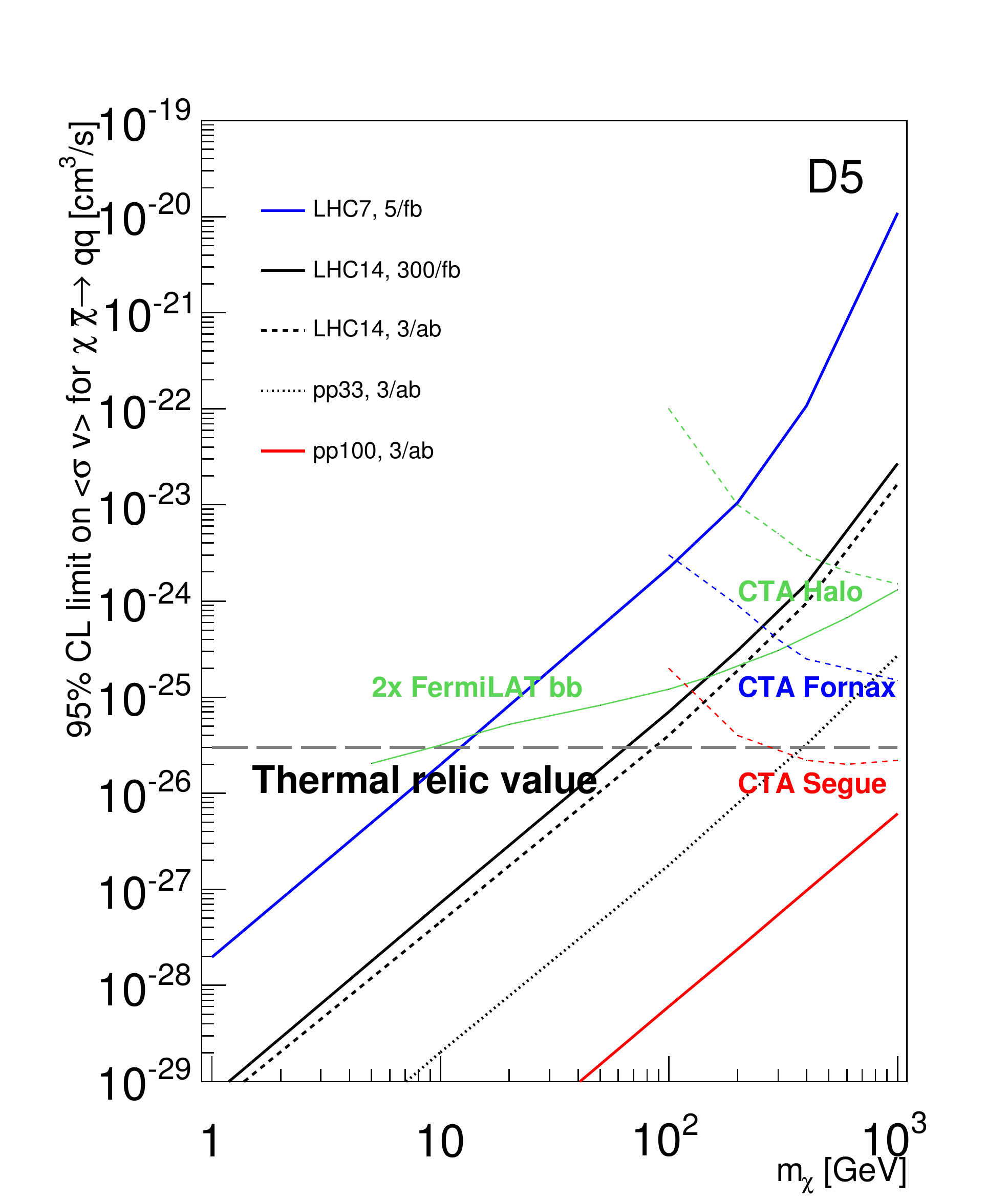}
\includegraphics[width=0.48\linewidth]{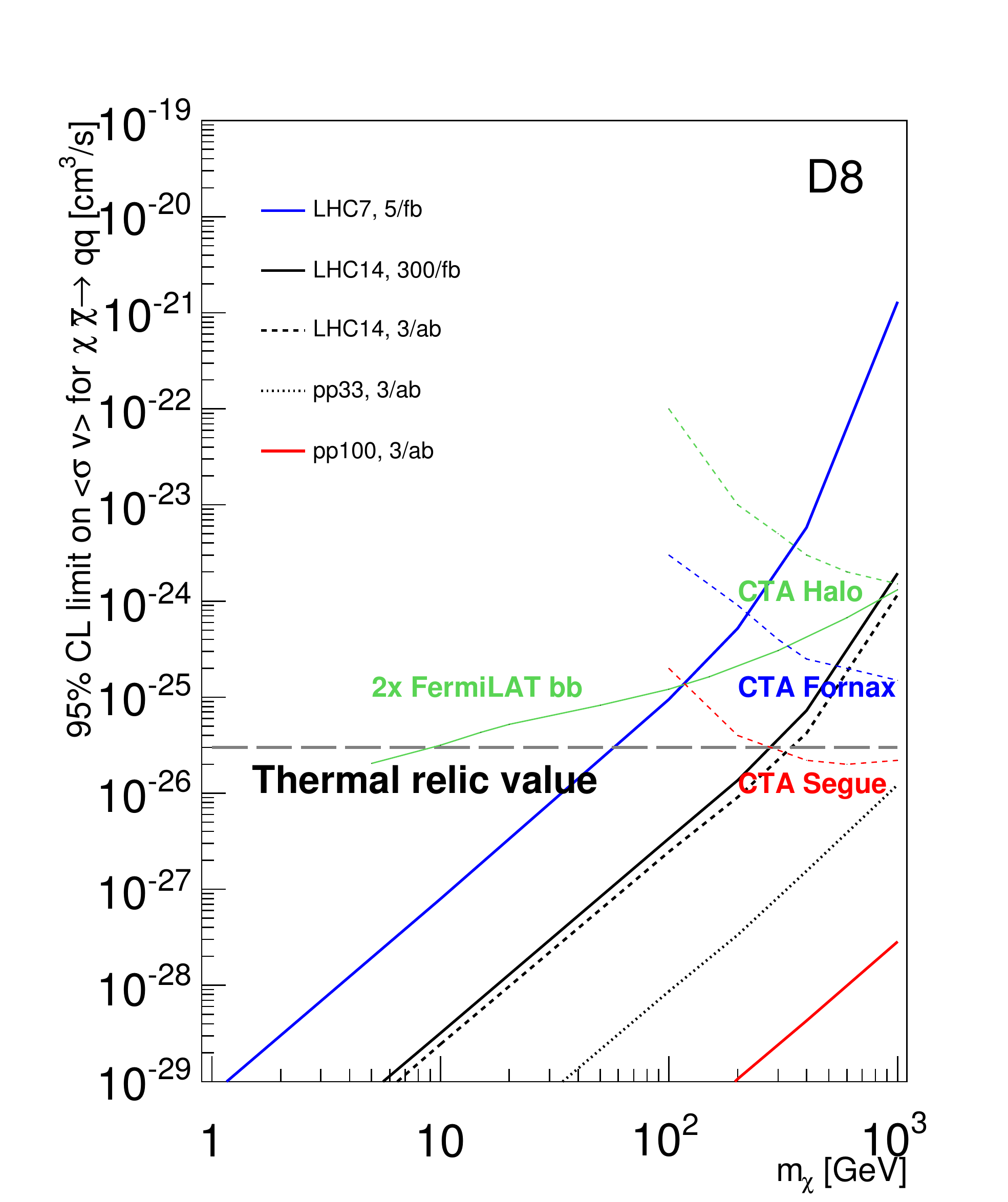}
\caption{Limits at 95\% CL on WIMP pair annihilation for different facilities using
  the D5 (left) or D8 (right) operator as a function of $m_\chi$.}
\label{fig:indirect}
\end{figure}

In Fig~\ref{fig:indirect}, we map to WIMP pair annihilation
cross-section limits. Our predictions are compared to Fermi-LAT limits
from a stacking analysis of Dwarf galaxies~\cite{fermi}, including a
factor of two to convert the Fermi-LAT limit from Majorana to Dirac
fermions, and to projected sensitivities of CTA~\cite{CTA}.

\section{Results for On-Shell Mediators}

The EFT approach is useful when the current facility does not have the
necessary center-of-mass energy to produce on-shell mediators.  The
next-generation facility, however, may have such power.  

In this section, we study the sensitivity of the proposed facilities
to a model in which the heavy mediator is a $Z'$ which couples to
$\chi\bar{\chi}$ as well as $q\bar{q}$~\cite{zprime,Frandsen:2012rk,Shoemaker:2011vi}. We generate
events as before, and measure the efficiency at parton level using
simulated events.

\begin{figure}[ht]
\includegraphics[width=0.48\linewidth]{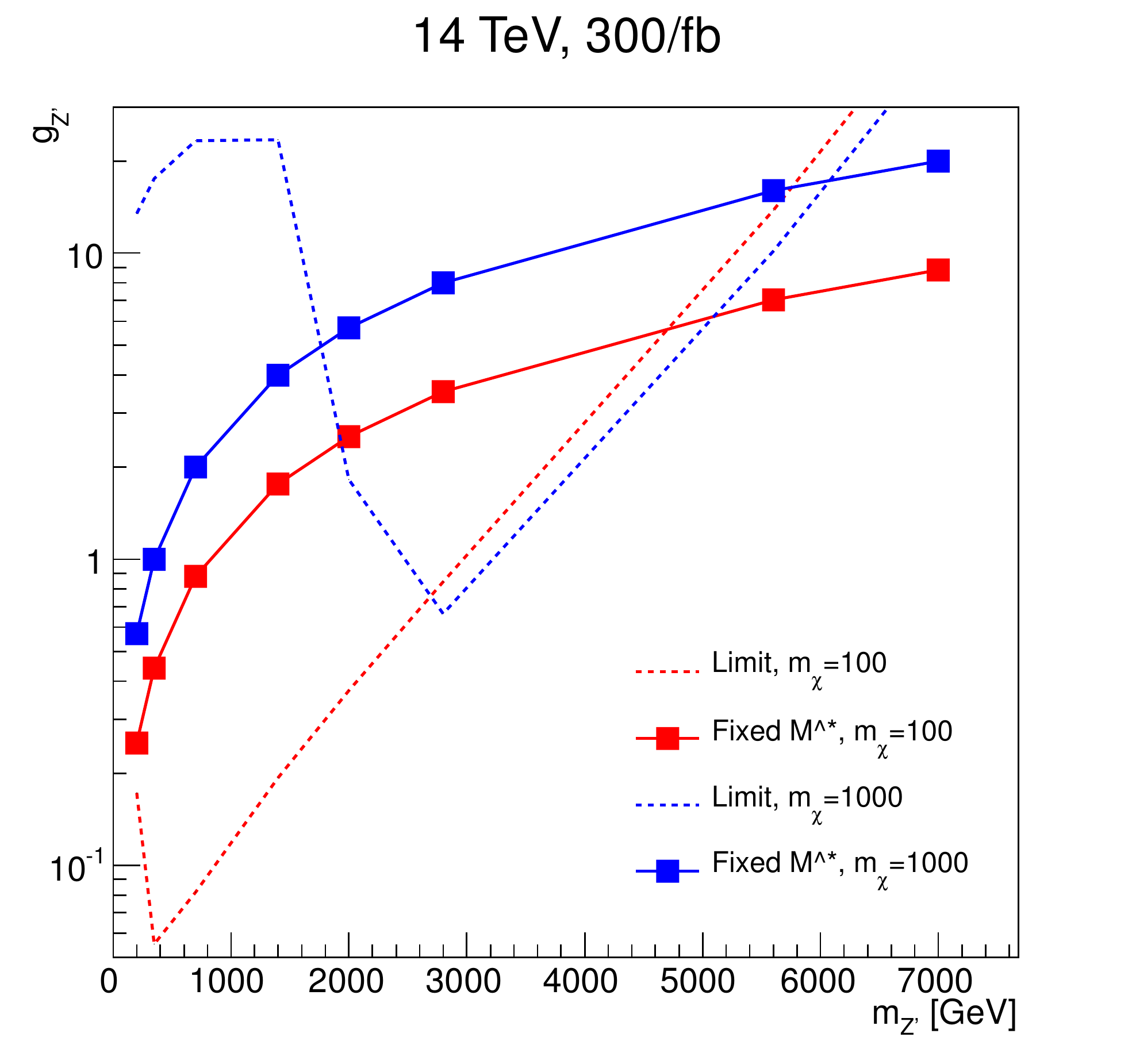}
\includegraphics[width=0.48\linewidth]{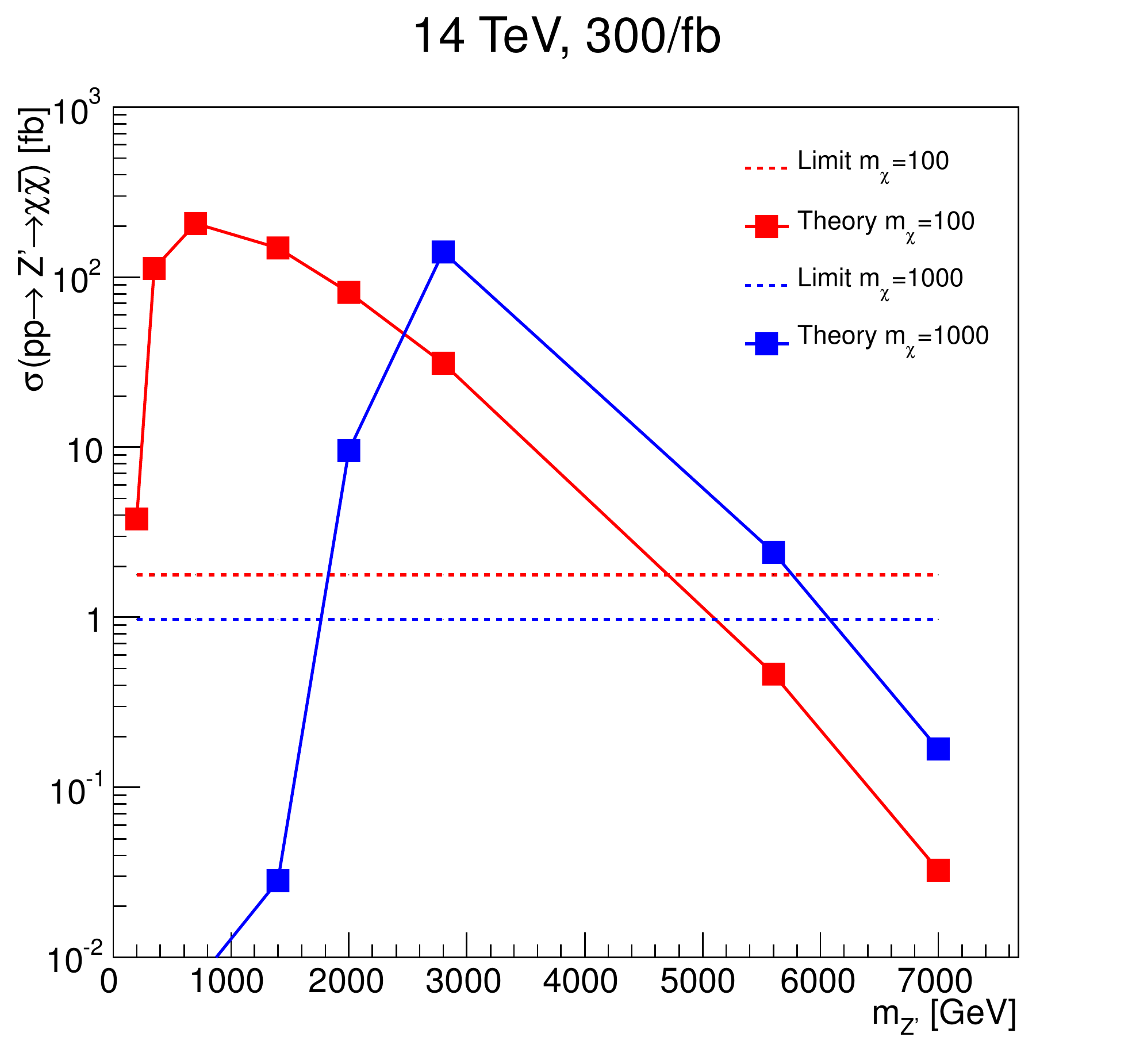}\\
\caption{ Sensitivity at $\sqrt{s}=14$
  TeV, $\mathcal{L}=300$ fb$^{-1}$ to a dark matter pairs produced through a real
  $Z'$ mediator. Left, expected limits on the coupling
  $g_{Z'}$ versus $Z'$ mass for two choices of $m_\chi$ for events
  with $\missET>550$ GeV; also shown are the values of $g_{Z'}$ which
  satisfy $g'/m_{Z'}=1/M_*$, where $M_*$ are limits from $\sqrt{s}=7$
  TeV, $\mathcal{L}=5$ fb$^{-1}$. Right, production cross section as a function of $Z'$ mass,
  compared to expected limits, where $g_{Z'}$ depends on $m_{Z'}$ as
  in the left pane.}
\label{fig:zp_14_300}
\end{figure}

The coupling of the $Z'$ is a free parameter in this theory, but
particularly interesting values are those which correspond to the
limit of previous facilities on $M_*$. That is, an EFT model of the
$Z'$ interaction has

\[ \frac{1}{M_*} = \frac{g_{Z'}}{M_{Z'}} \]

fixing the relationship between $g_{Z'}$ and
$M_{Z'}$. Figure~\ref{fig:zp_14_300} shows the expected limits on the
$Z'$ model at a facility with $\sqrt{s}=14$ TeV and $\mathcal{L}=300$
fb$^{-1}$, in terms of the cross section $\sigma(pp\rightarrow
Z'\rightarrow \chi\bar{\chi} j$ for $\missET>550$ GeV and in terms of
$g_{Z'}$. The $g'$ expected limits can be compared to the curve with
$g_{Z'} = \frac{M_{Z'}}{M_*}$; the cross-section limits can be
compared to the predicted cross section assuming $g_{Z'} = \frac{M_{Z'}}{M_*}$.

\begin{figure}[th]
\includegraphics[width=0.48\linewidth]{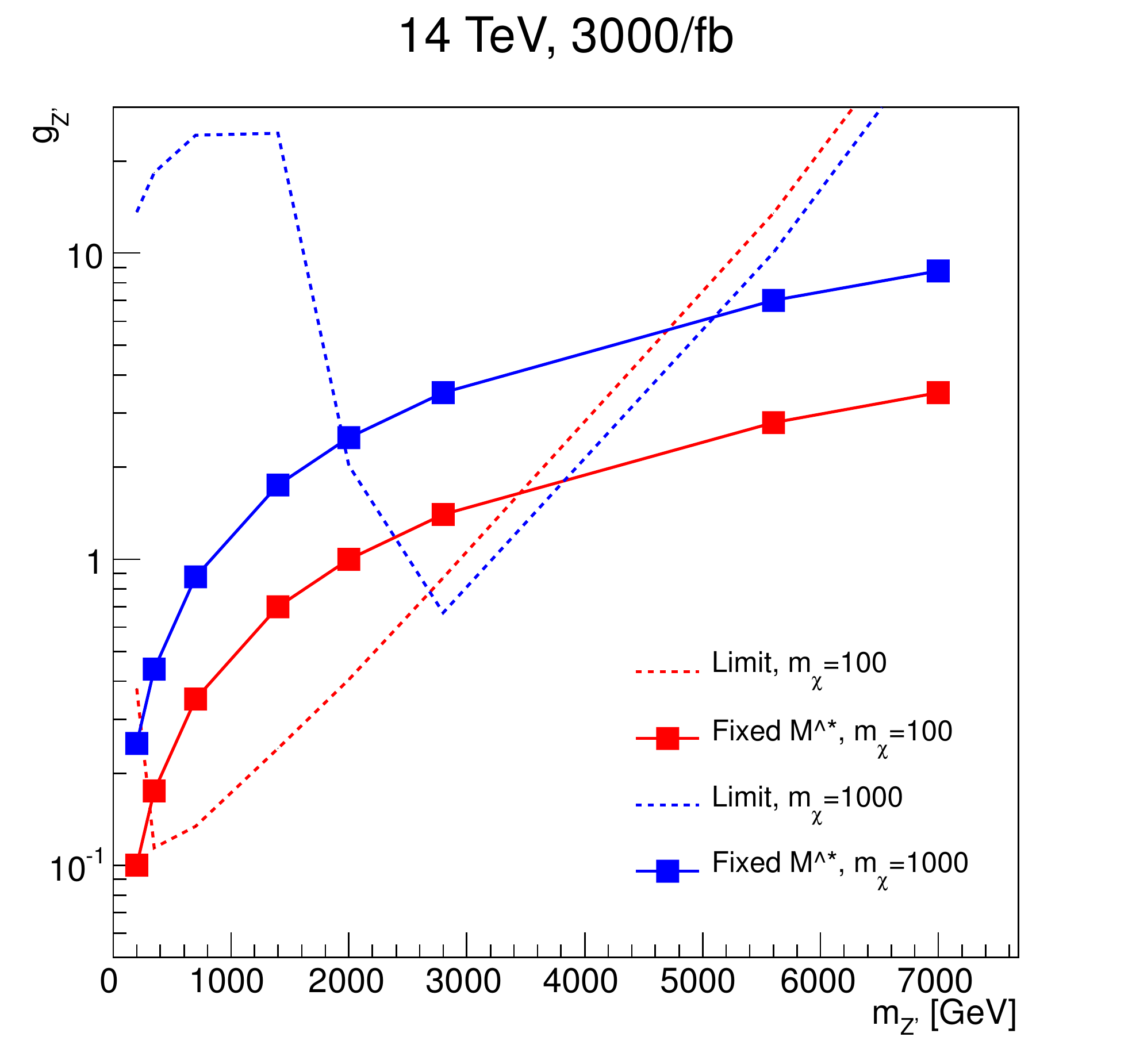}
\includegraphics[width=0.48\linewidth]{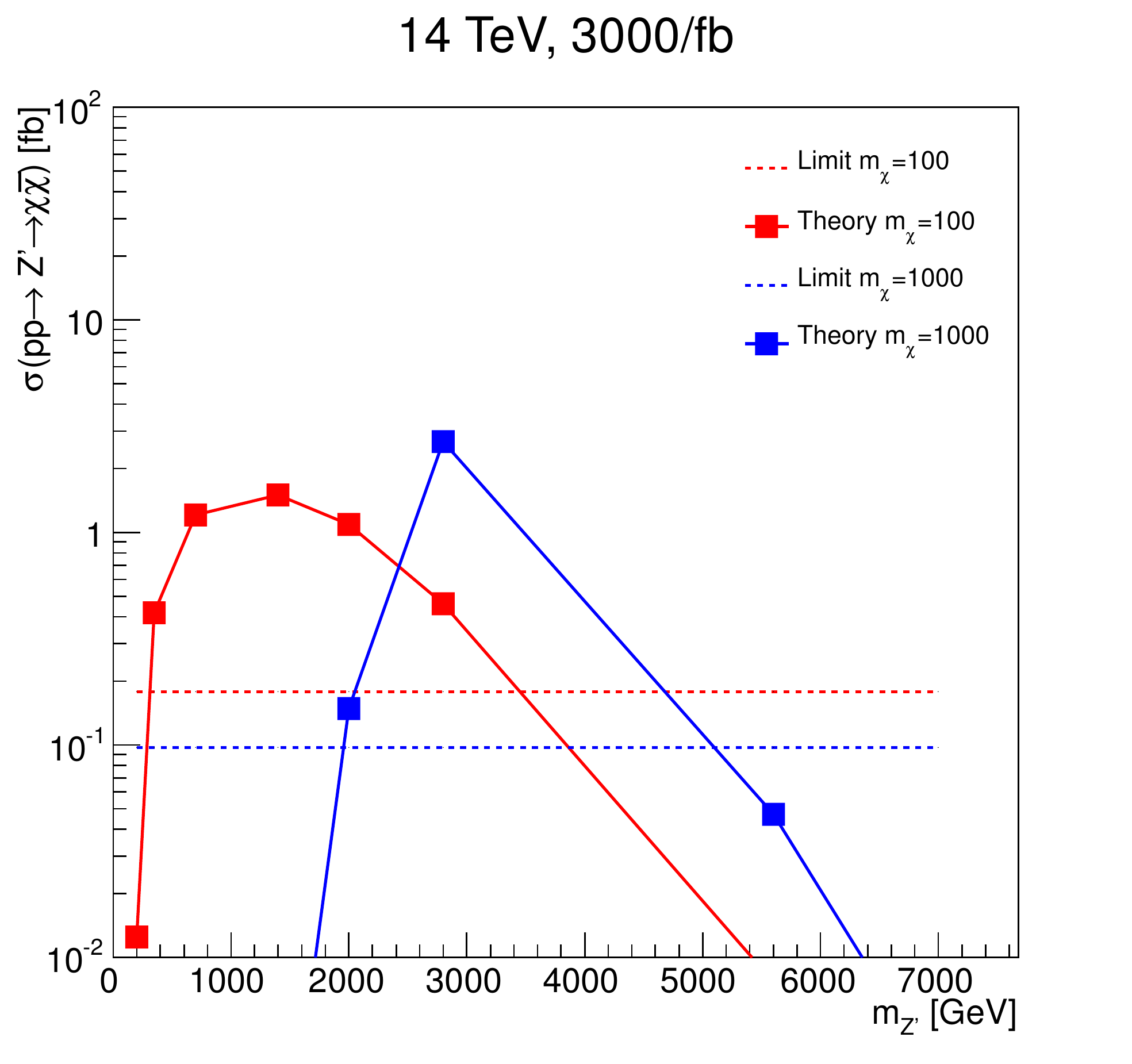}\\
\caption{ Sensitivity at $\sqrt{s}=14$
  TeV, $\mathcal{L}=3000$ fb$^{-1}$ to a dark matter pairs produced through a real
  $Z'$ mediator. Left, expected limits on the coupling
  $g_{Z'}$ versus $Z'$ mass for two choices of $m_\chi$ for events
  with $\missET>1100$ GeV; also shown are the values of $g_{Z'}$ which
  satisfy $g'/m_{Z'}=1/M_*$, where $M_*$ are limits from $\sqrt{s}=14$
  TeV, $\mathcal{L}=300$ fb$^{-1}$. Right, production cross section as a function of $Z'$ mass,
  compared to expected limits, where $g_{Z'}$ depends on $m_{Z'}$ as
  in the left pane.}
\label{fig:zp_14_3000}
\end{figure}

\begin{figure}[th]
\includegraphics[width=0.48\linewidth]{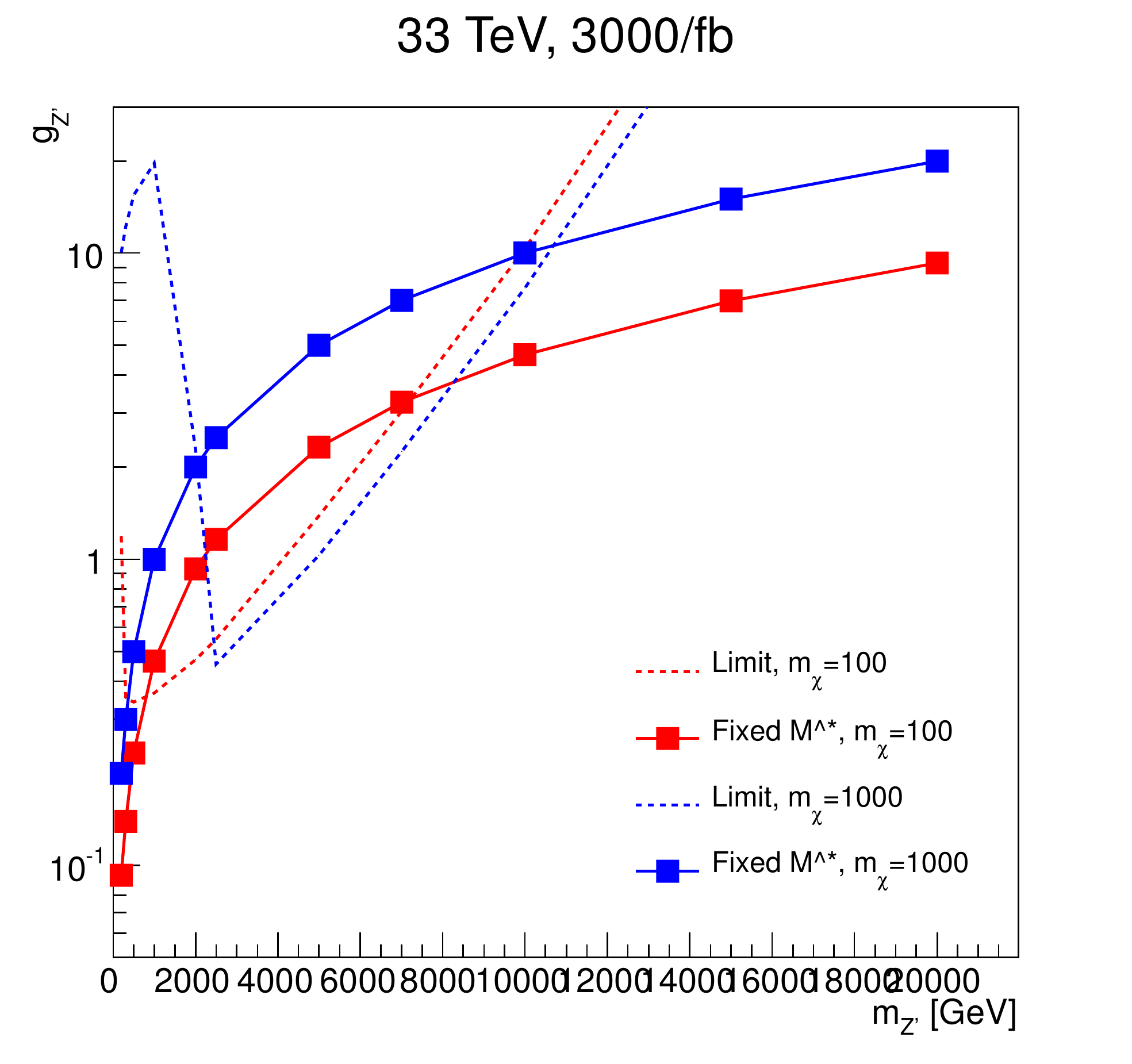}
\includegraphics[width=0.48\linewidth]{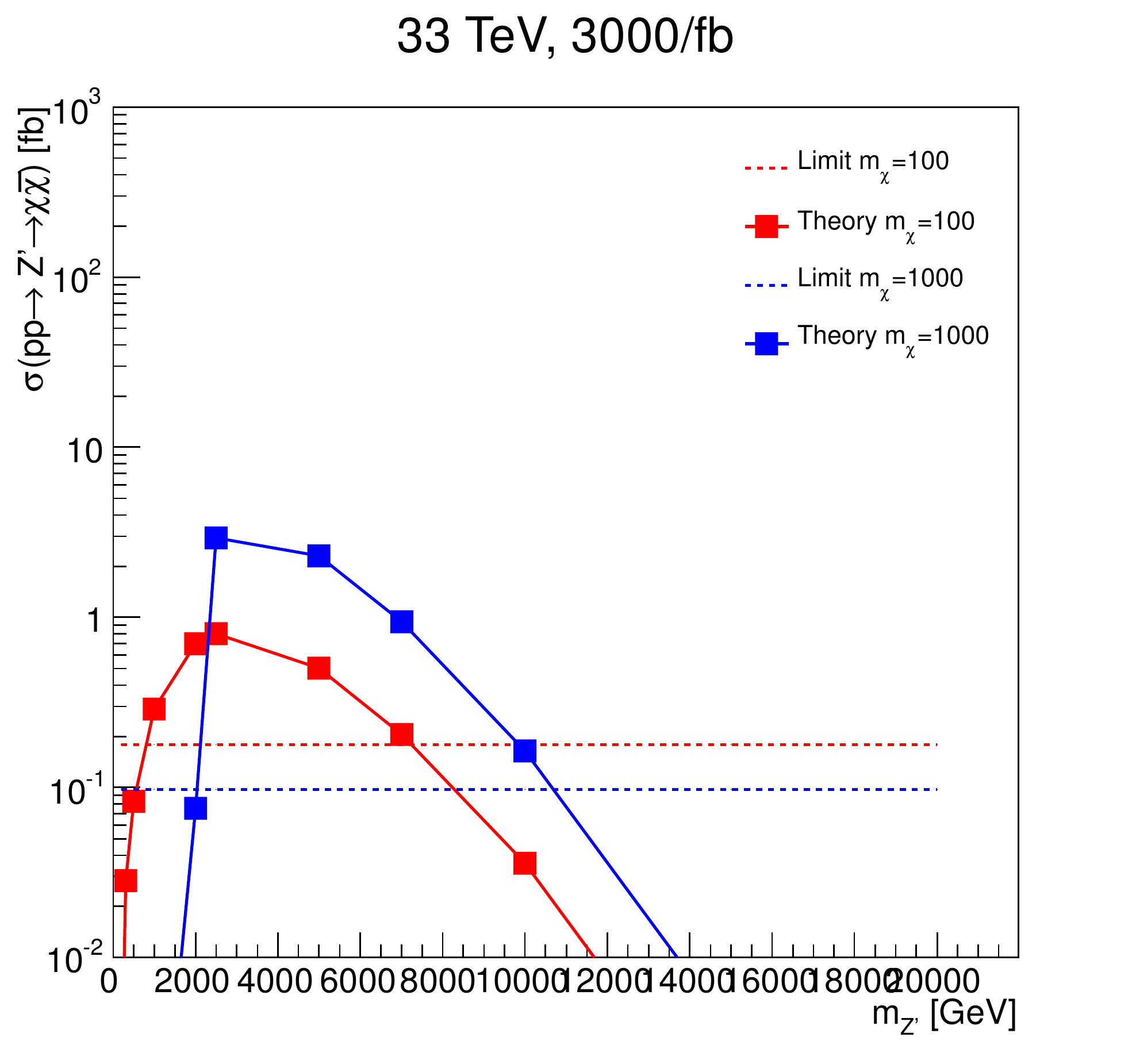}\\
\caption{ Sensitivity at $\sqrt{s}=33$
  TeV, $\mathcal{L}=3000$ fb$^{-1}$ to a dark matter pairs produced through a real
  $Z'$ mediator. Left, expected limits on the coupling
  $g_{Z'}$ versus $Z'$ mass for two choices of $m_\chi$ for events
  with $\missET>2750$ GeV; also shown are the values of $g_{Z'}$ which
  satisfy $g'/m_{Z'}=1/M_*$, where $M_*$ are limits from $\sqrt{s}=14$
  TeV, $\mathcal{L}=3000$ fb$^{-1}$. Right, production cross section as a function of $Z'$ mass,
  compared to expected limits, where $g_{Z'}$ depends on $m_{Z'}$ as
  in the left pane.}
\label{fig:zp_33_3000}
\end{figure}

\begin{figure}[ht]
\includegraphics[width=0.48\linewidth]{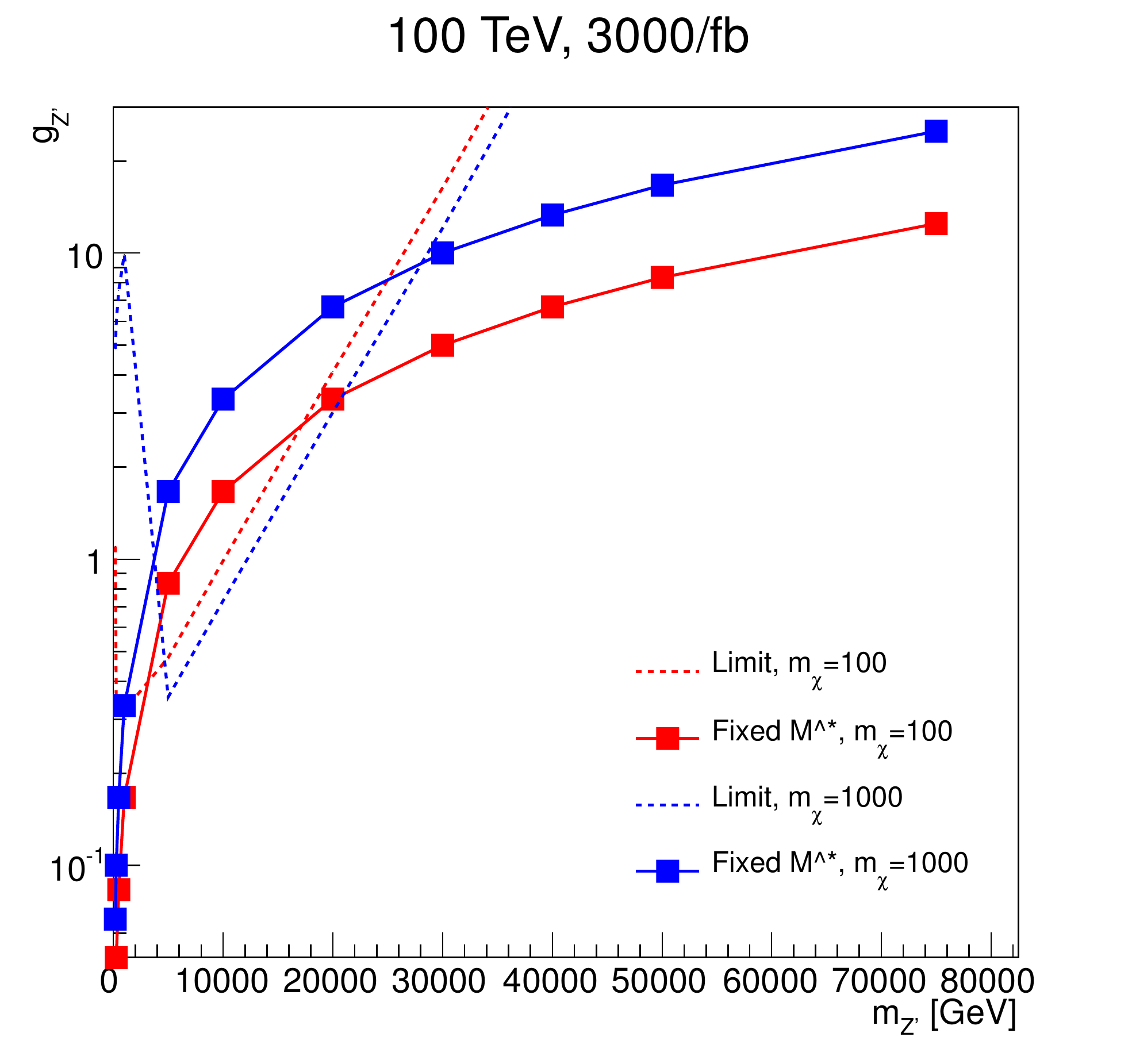}
\includegraphics[width=0.48\linewidth]{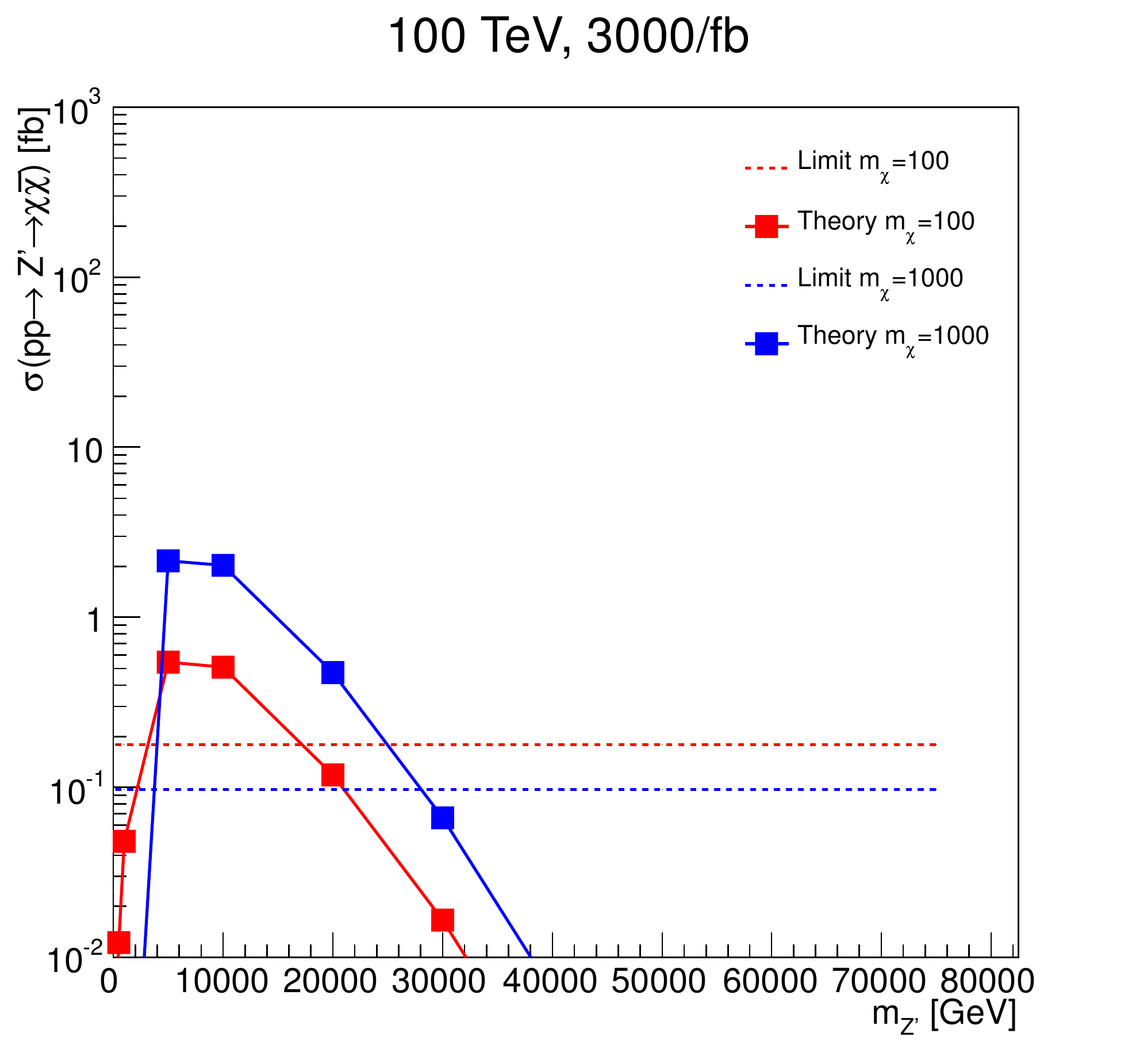}\\
\caption{ Sensitivity at $\sqrt{s}=100$
  TeV, $\mathcal{L}=3000$ fb$^{-1}$ to a dark matter pairs produced through a real
  $Z'$ mediator.  Top, expected limits on the coupling
  $g_{Z'}$ versus $Z'$ mass for two choices of $m_\chi$ for events
  with $\missET>5500$ GeV; also shown are the values of $g_{Z'}$ which
  satisfy $g'/m_{Z'}=1/M_*$, where $M_*$ are limits from $\sqrt{s}=33$
  TeV, $\mathcal{L}=3000$ fb$^{-1}$. Bottom, production cross section as a function of $Z'$ mass,
  compared to expected limits, where $g_{Z'}$ depends on $m_{Z'}$ as
  in the top pane.}
\label{fig:zp_100_3000}
\end{figure}

Similar results for other facilities are shown in
Figures~\ref{fig:zp_14_3000}, ~\ref{fig:zp_33_3000}, and ~\ref{fig:zp_100_3000}.

\subsection{Acknowledgements}

We acknowledge useful conversations with Roni Harnik and Patrick Fox.
DW and NZ are supported by grants from the Department of Energy
Office of Science and by the Alfred P. Sloan Foundation. 
The research of TMPT is supported in part by NSF
grant PHY-0970171 and by the University of California, Irvine through a Chancellor's fellowship.

\end{document}